\documentclass[aps,twocolumn,groupedaddress,showpacs,floats,amssymb]{revtex4}
\usepackage[dvips]{graphicx}
\usepackage{bm}

\begin{document}

\noindent {\bf Comment on ``Direct Mapping of the Finite Temperature Phase Diagram of Strongly Correlated Quantum Models"}\vspace{2mm}\\

In their Letter~\cite{Zhou09}, Zhou, Kato, Kawashima, and Trivedi claim that finite-temperature critical points of strongly correlated
quantum models emulated by optical lattice experiments can generically be
deduced from kinks in the derivative of the density profile of atoms in the trap
with respect to the external potential, $\kappa = -dn(r)/dV(r)$.
In this comment we demonstrate that the authors failed to achieve their goal: 
to show that under realistic experimental conditions critical densities 
$n_c(T,U)$ can be extracted from density profiles with controllable accuracy.

When illustrating their proposal with numerical data (in Figs.~4 and 5), the authors  take it for granted that (i) sharp features in $\kappa $ only come from the critical behavior (Fig.~4), and (ii) critical behavior in a trapped system necessarily results in a cusp in $\kappa$ (Fig.~5).
Both assumptions are wrong, and this invalidates the first-principles component of the work.

(i) Sharp features in $\kappa$ do not {\it necessarily} originate from critical fluctuations. A relevant counterexample would be superfluid helium in a Dewar
with the density gradient sharply peaked at the wall. The only direct simulation performed 
by the authors and presented in Fig.~4 falls in this category. 
Both sharp features correspond to extremely large gradients 
of particles (near the trap perimeter) and holes (near the Mott insulator phase),
when the particle/hole densities change by $\sim 100\%$ 
over one lattice spacing in the radial direction! 
In this case it is {\it fundamentally} impossible to extract the critical concentration of particles, $n_c(T,U)$,  or holes, $1-n_c(T,U)$,
with controllable accuracy since the critical region is simply absent.

(ii) The four illustrations presented in Fig.~5 are based on the 
local density approximation (LDA) reconstruction, 
not the results of simulations in the trap, and are thus misleading. 
In all cases the density changes by $\sim 100\%$ at a distance of 
just a few lattice spacings, meaning that the size of the critical 
region is limited, and features in $\kappa$ 
must be dramatically rounded. The effect of finite-size
rounding of critical singularities is routinely seen in 
Monte Carlo simulations but the authors do not address this problem at all.
By the very nature of  second-order phase transitions LDA inevitably fails because of a 
divergent correlation radius in the vicinity of a critical point, 
and this is precisely why the sharp features get rounded.
The reader should not be misled by the sharp features in $\kappa$
because pronounced kinks in Fig.~5 are hypothetical and 
do not represent the actual behavior of $\kappa$ in a trapped system! 

\begin{figure}[htb]
\includegraphics[bb=0 110 600 700, angle=-90, width=0.68\columnwidth]{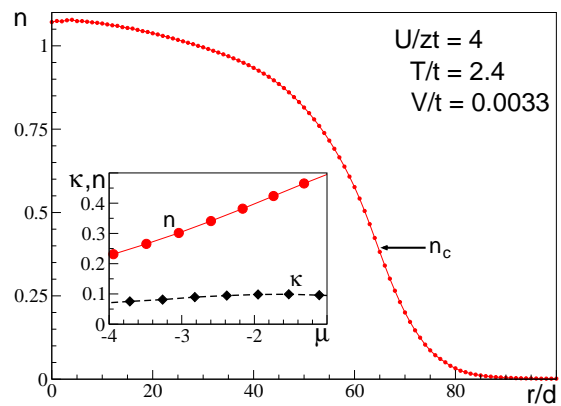}
\caption{Density profile and its derivative with respect to the 
external potential (inset) for parameters of Fig. 5b in Ref.~\cite{Zhou09}. Our unit is the hopping amplitude $t$. Errors are on the order of the point size, or smaller.
} \label{Fig1}
\end{figure}

To show how severe the rounding of critical singularities in a trapped 
system is, and how it makes the precise determination of critical parameters 
from $\kappa$ nearly impossible, we performed simulations of the system specified in 
Fig.~5(B). The first-principles data are presented in Fig.~\ref{Fig1}. 

In sharp contrast with  the LDA plots of Ref.~\cite{Zhou09}, we observe no 
features for the published parameters that would allow one to define errorbars through the full width at 
half maximum (or any characteristic interval) of the experimental curve. This brings us 
to the above-formulated conclusion. It leaves the question open whether it is feasible 
to extract critical parameters from the density~\cite{LDA_violations}.

This work was supported by the Swiss National Science Foundation,  the National Science Foundation under Grant PHY-0653183, and a grant from the Army Research Office with funding from the DARPA OLE program.
Simulations were performed on the Brutus cluster at ETH Zurich and use was made of the ALPS libraries for the error evaluation~\cite{ALPS}.
\vspace{2mm}\\
{\small \noindent   Lode Pollet$^1$, Nikolay Prokof'ev$^2$, and Boris Svistunov$^2$,\\
$^1$Physics Department, Harvard University, Cambridge, MA 02138.\\
$^2$Department of Physics,  University of Massachusetts, Amherst, MA 01003.}
\bigskip



\begin{thebibliography}{99}
\bibitem{10_percent} An accuracy of $\pm10\%$ was the standard set by the DAPRA OLE program. 
Densities outside the  range ($n=0.2$ to $n=0.5$) shown in the inset of  Fig.~\ref{Fig1} are thus irrelevant.
\end{thebibliography}
\end{document}